\documentclass[aps,pre,twocolumn,preprintnumbers,amsmath,amssymb,superscriptaddress,showkeys]{revtex4-1}

\usepackage{hyperref}
\usepackage{amssymb}
\usepackage{graphicx}
\usepackage{amsmath}
\usepackage{amsfonts}
\usepackage[english]{babel}
\usepackage{color}

\begin{document}
\author{Isabelle Schneider}
\affiliation{Institut f{\"u}r Mathematik, Freie Universit\"at Berlin, Arnimallee 7, 14195 Berlin, Germany}
\author{Marie Kapeller}
\affiliation{Institut f{\"u}r Theoretische Physik, Technische Universit\"at Berlin, Hardenbergstra\ss{}e 36, 10623 Berlin, Germany}
\author{Sarah Loos}
\affiliation{Institut f{\"u}r Theoretische Physik, Technische Universit\"at Berlin, Hardenbergstra\ss{}e 36, 10623 Berlin, Germany}
\author{Anna Zakharova}
\affiliation{Institut f{\"u}r Theoretische Physik, Technische Universit\"at Berlin, Hardenbergstra\ss{}e 36, 10623 Berlin, Germany}
\author{Bernold Fiedler}
\affiliation{Institut f{\"u}r Mathematik, Freie Universit\"at Berlin, Arnimallee 7, 14195 Berlin, Germany}
\author{Eckehard Sch{\"o}ll}
\affiliation{Institut f{\"u}r Theoretische Physik, Technische Universit\"at Berlin, Hardenbergstra\ss{}e 36, 10623 Berlin, Germany}

\email[corresponding author:] {schoell@physik.tu-berlin.de}

\title{\bf{Stable and transient multi-cluster oscillation death in nonlocally coupled networks}}

\begin{abstract}
In a network of nonlocally coupled Stuart-Landau oscillators with symmetry-breaking coupling, we study numerically, and explain analytically, a family of inhomogeneous steady states (oscillation death). They exhibit multi-cluster patterns, depending on the cluster distribution prescribed by the initial conditions. Besides stable oscillation death, we also find a regime of long transients asymptotically approaching synchronized oscillations.
To explain these phenomena analytically in dependence on the coupling range and the coupling strength,
we first use a mean-field approximation which works well for large coupling ranges but fails for coupling ranges which are small compared to the cluster size. Going beyond standard mean-field theory, we predict the boundaries of the different stability regimes as well as the transient times analytically in excellent agreement with numerical results.
\end{abstract}


\date{\today}
\maketitle

\section{Introduction}\label{i}

Coupled nonlinear systems exhibit a plethora of different collective behavior such as partial or cluster synchronization \cite{PIK01,BOC02,DAH12,DO12,PEC14,POE15}, spatio-temporal patterns under delayed feedback control \cite{FIE09,SCH13b,SCH14d}, and chimera states which consist of spatially coexisting domains of synchronized and unsynchronized dynamics \cite{KUR02a,ABR04,PAN15}. Recently, special attention has been paid to the different types of oscillation quenching, amplitude death and oscillation death \cite{ZOU09a,PRA10,KOS10,KOS13}. Amplitude death and oscillation death differ in the mechanisms by which they are induced:
In coupled oscillator networks, the coupling between oscillators can stabilize an already existing homogeneous steady state which is unstable in the absence of coupling.
This phenomenon is called \emph{amplitude death}. 
In contrast, \emph{oscillation death} requires the breaking of the system's symmetry:
Here inhomogeneous steady states are newly created as well as stabilized through a symmetry-breaking coupling between the oscillators.
A combination of these two phenomena of spontaneous symmetry-breaking, chimera states and oscillation death, has recently been
found in a paradigmatic model of nonlocally coupled supercritical Hopf normal forms (Stuart-Landau oscillators) \cite{ZAK14,ZAK15b}. The resulting patterns of coexisting coherent and incoherent inhomogeneous steady states occur in form of clusters of different size, and have been named {\em chimera death}. Similar chimera death patterns have recently also been observed with global (all-to-all) coupling \cite{BAN15,PRE15}. 

Oscillation death has been observed experimentally in many different systems, such as chemical reactors \cite{DOL88}, chemical oscillators \cite{CRO89}, chemical droplets \cite{TOI08}, electronic circuits \cite{HEI10}, or thermokinetic oscillators \cite{ZEY01}.
There exist various biological  applications, e.g.~neural networks \cite{CUR10}, genetic oscillators \cite{KOS10a}, calcium oscillators \cite{TSA06} or stem cell differentiation \cite{SUZ11} where it has been proposed as a basic mechanism for morphogenesis and cellular differentiation.
Consequently, there has also been theoretical interest in oscillation death and it has been shown to exist for special time-delayed \cite{ZAK13a} and repulsive \cite{HEN13, NAN14} types of coupling as well as coupling through conjugate or dissimilar variables \cite{LIU12c, ZOU13, KAR14}.
However, research has mostly focussed on small numbers of coupled oscillators in networks with local or global coupling.

The influence of nonlocal coupling on the onset of oscillation death and on the emergence of multi-cluster patterns in larger networks has not been investigated systematically, so far.
It is possible to use the coupling range $P$ as a control parameter, e.g., in a bidirectionally coupled ring of $N$ nodes. Thus one can interpolate between the well-studied limit cases of local (nearest neighbor, $P = 1$) and global (all-to-all, $P = \lfloor N/2\rfloor $) coupling. 

In our present study, we analyze the transition from completely synchronized oscillations to multi-cluster inhomogeneous steady state
patterns (oscillation death) in networks of nonlocally coupled oscillators with symmetry-breaking coupling. We choose initial conditions which favor the emergence of spatially coherent clusters, in contrast to \cite{ZAK14}, where hybrid patterns with 
coexisting domains of spatially coherent and spatially incoherent steady states (chimera death) were analyzed.
We investigate the effects of several factors on the occurrence of different oscillation death states both numerically and analytically.
First, we use nonlocal coupling and thus cover the whole range between the limit cases of local and global coupling.
Second, we also vary the initial conditions, specified by cluster distributions, to find different coexisting spatial patterns for oscillation death. Furthermore, we also vary the coupling strength, and combine all three variations.
Beside the regions of stability, we also find parameter regimes where the oscillation death state is transient, but persists for a long time. Due to the coherent nature of the oscillation death patterns, we are able to predict analytically and with high precision the boundaries between different stability regimes.

This paper is organized as follows:
In section~\ref{ii} we introduce our model of nonlocally coupled oscillators and briefly summarize previous work on two coupled oscillators \cite{ZAK13a}.
Our approach is numerical in the first step, and we observe two different types of oscillation death: transient as well as asymptotically stable.
In section~\ref{iii} we present our numerical findings.
In the next step, we explain our findings analytically, such as boundaries in the parameter space for the occurrence of the different patterns of oscillation death.
In section~\ref{iv}, we use a mean-field ansatz to obtain a first approximation of the stability boundaries.
In section~\ref{v} we go beyond the mean-field approximation to describe the boundaries analytically in excellent agreement with the numerical results. 
In section~\ref{vi} we calculate the transient times of transient oscillation death analytically, thereby gaining insight into the mechanism. In section~\ref{vii} we draw some conclusions.  

\section{Model}\label{ii}

In the present study, we investigate a network of coupled Stuart-Landau oscillators.
The equation of the single Stuart-Landau oscillator, which is a truncated generic expansion of a system near supercritical Hopf bifurcation in center manifold coordinates, is given by
\begin{equation} \label{eq:StuLa}
\dot{z} = f(z) \equiv \left(\lambda + i\omega- |z|^2\right)z.
\end{equation}
Here $z = re^{i\phi} = x + i y \in \mathbb{C}$ denotes the phase space, and $\lambda \in \mathbb{R}$ can be viewed as a bifurcation parameter. 
A Hopf bifurcation occurs at $\lambda=0$.
For $\lambda>0$, a stable periodic orbit with radius $r_0 =\sqrt{\lambda}$ appears, which oscillates with angular velocity $\omega \in \mathbb{R}$.
This periodic orbit $z(t)=r_0 e^{i \omega t}$ is harmonic and rotationally symmetric. 
In mathematical terms, we say that the single Stuart-Landau oscillator is $S^1$-equivariant, i.e.~$z(t)$ solves $\dot{z}=f(z)$ if and only if $e^{i\theta}z(t)$ solves $\dot{z}=f(z)$, for any fixed phase angle $\theta\in [0,2\pi]= S^1$.
Indeed $f$ commutes with the $S^1$-action of $\theta$ on $\mathbb{C}^N$: $f(e^{i\theta}z)=e^{i\theta}f(z)$ for all $\theta$ and $z$.

We arrange $N$ Stuart-Landau oscillators $z_j$, $j = 1,2,\dots,N$, with $N$ even, as a ring and couple them as follows:
\begin{equation}\label{eq:StuLaRing}
\dot{z_j}=f(z_j)+ \frac{\sigma}{2P}\sum_{k=j-P}^{j+P} \left(\mbox{Re}\,z_k - \mbox{Re}\,z_j\right)  
\end{equation}
where all indices are modulo $N$.
The \emph{coupling strength} is described by the real parameter $\sigma >0$.
Each oscillator $z_j$ is coupled to the oscillators $z_{j-1},\dots,z_{j-P}$ as well as to the oscillators $z_{j+1},\dots,z_{j+P}$, i.e., to its $P$ nearest neighbors in each direction of the ring, respectively.
The \emph{coupling range} is given by $P$, where $P$ corresponds to the number of coupling neighbors in each direction on a ring. 
The coupling is normalized by the number of existing links, which is $2P$.

In this publication, we focus on the case of \emph{nonlocal} coupling. 
Note, however, that our results also hold true for the case of local coupling, which is given by $P=1$, and in the limit of large $N$ for global coupling, which is given approximately by $P=N/2$. Strictly speaking, the limit of global coupling exists only for odd $N$ and $P=(N-1)/2$, since for even $N$ the antipodal link would appear twice.

Here we consider coupling only in the real part $x_j$ of the variable $z_j$, which breaks the $S^1$-equivariance of the single Stuart-Landau oscillator: 
$Re(e^{i\theta}z_j) = e^{i\theta}Re(z_j)$ is only true for $\theta =0$ or $\theta =\pi$.
Symmetry-breaking  of $S^1$-equivariance is a necessary condition for the existence of isolated nontrivial steady states with $z_j \ne 0 $ and thus for oscillation death.

Note, however, that the rotationally symmetric periodic orbit of the single Stuart-Landau oscillator is preserved by the coupling.
It exists for all coupling strengths $\sigma$ and can be seen if all oscillators are in synchrony.

\begin{figure}[tbp]
\begin{center}
\includegraphics[width=\linewidth]{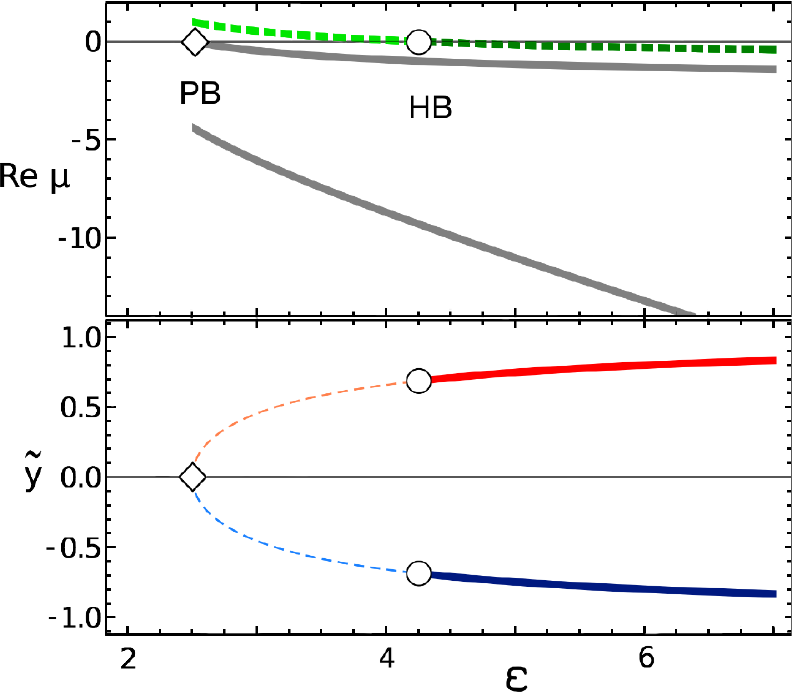}
\caption{Top: Real part of the complex conjugate eigenvalues $\mu_{1,2}$ (green dashed) and of the real eigenvalues $\mu_3$ and $\mu_4$ (gray solid) as a function of the coupling strength $\varepsilon$ for 2 coupled oscillators. Bottom: Imaginary variable $\tilde{y}$ of the inhomogeneous steady states versus $\varepsilon$. Parameters: $\lambda=1$, $\omega=2$. There is a pitchfork bifurcation (PB) at $\varepsilon=2.5$, indicated by a diamond, and secondary Hopf bifurcations (HB) at $\varepsilon=4.25$, indicated by circles.
}
\label{triplett}
\end{center}
\end{figure}

In the present study, we restrict our parameters to the interesting case $0<\lambda<\omega$.
The following results, which will be needed later, have been derived in \cite{ZAK13a} for a system of only two coupled Stuart-Landau oscillators:
\begin{equation}\label{2osci}
\begin{split}
\dot{z}_{1} &= f(z_1)+\varepsilon \left(\mbox{Re}\, z_{2}-\mbox{Re}\, z_1\right)\\
\dot{z}_{2} &= f(z_2)+\varepsilon \left(\mbox{Re}\, z_{1}-\mbox{Re}\, z_2\right),
\end{split}
\end{equation}
with a coupling parameter $\varepsilon>0$.
In the \emph{in-phase subspace} $z_1\equiv z_2$, this system simplifies to $\dot{z} = f(z)$, i.e., to a single Stuart-Landau oscillator.
In the \emph{anti-phase subspace} $z_1\equiv -z_2$, this system simplifies to $\dot{z} = f(z)-2 \varepsilon \,\mbox{Re}\, z$.
The system \eqref{2osci} exhibits a stable synchronous periodic orbit for all coupling parameters $\varepsilon$.
There also exists a trivial homogeneous steady state $z_1\equiv z_2\equiv0$.
Linearization at zero reveals that this state is always unstable.
Supercritical pitchfork bifurcation occurs at 
\begin{equation}
\varepsilon_{\scriptsize{P}}=\tfrac{1}{2}\left(\lambda+\tfrac{\omega^2}{\lambda}\right).
\end{equation} 
At this pitchfork bifurcation, two inhomogeneous steady state solutions $(\tilde{x}_{1,2},\tilde{y}_{1,2})$ 
\begin{align}
\tilde{x}_{1,2} &= \pm\sqrt{\lambda-\varepsilon +\sqrt{\varepsilon^2-\omega^2}-\tilde{y}^2} \label{x}\\
\tilde{y}_{1,2} &= \mp \sqrt{\left(\lambda\varepsilon - \omega^2 + \lambda\sqrt{\varepsilon^2 - \omega^2}\right)/(2\varepsilon)}\label{y}
\end{align}
emanate for $\varepsilon>\varepsilon_P$. They correspond to antiphase states $\tilde{z}_1=-\tilde{z}_2$.
The radius $\tilde{r}$ of the inhomogeneous steady state is given explicitly by
\begin{equation}
\tilde{r}^2 =\lambda - \varepsilon + \sqrt{\varepsilon^2 - \omega^2}.
\end{equation}
Linearizing at those steady states, we find the eigenvalues $\eta$.
There exists a pair of complex conjugate eigenvalues in direction of the in-phase subspace,
\begin{equation}
\mu_{1,2} = \lambda - 2 \tilde{r}^2 \pm  i\sqrt{\omega^2 - \tilde{r}^4},
\end{equation}
and there exist two real and distinct eigenvalues in direction of the anti-phase subspace,
\begin{equation}
\mu_{3,4} = -\lambda+\varepsilon - 2 \sqrt{\varepsilon^2-\omega^2}\pm(\lambda-\varepsilon).
\end{equation}
In particular we encounter secondary Hopf bifurcations at 
\begin{equation}
\varepsilon_{H\!B} = \tfrac{1}{4}\left(\lambda + 4\tfrac{\omega^2}{\lambda}\right),
\end{equation}
where the inhomogeneous steady states are stabilized, thus marking the onset of oscillation death.
The secondary eigenvalues $\mu_{1,2}$ remain complex conjugate eigenvalues, with relatively small real part.
In contrast, $\mu_{3,4}$ give two distinct real and negative eigenvalues.
See Fig.~\ref{triplett} for a visualization of the eigenvalues and inhomogeneous steady states.

\section{Numerical simulations}\label{iii}

As a first step in the analysis of the system, we solve Eqs.~\eqref{eq:StuLaRing} numerically. 
We vary the coupling range $P$ from $0$ (no coupling) to $N/2$ (global coupling) in steps of 1. 
The intermediate range $0<P<N/2$ is called \emph{nonlocal coupling} and our main interest lies here. 
Additionally, we also increase the coupling strength $\sigma$ from $\sigma=1$ to $30$ with stepsize 1.
The simulations are run no longer than time $t=4000$.
The other parameters are fixed to $N=100$, $\lambda=1$, and $\omega=2$.
As initial conditions, we use \emph{anti-phase clusters} of size $n$:
In a first step, all oscillators  $z_1, \dots, z_n$ of the first cluster are set to the values $(x_j,y_j)=(-1,+1)$, which we call the \emph{upper branch} since $y_j$ is close to the upper steady state $\tilde{y}$ (cf. bottom panel of Fig.~\ref{triplett}).
In contrast, the oscillators $z_{n+1},\dots, z_{2n}$ are set to values on the \emph{lower branch} with $(x_j,y_j)=-(-1,+1)=(+1,-1)$.
Then oscillators $z_{2n+1}, \dots, z_{3 n}$ are again on the upper branch and so forth.  
In a second step, a random number, drawn form an underlying Gaussian distribution with zero mean and variance $0.1$, is added to the $y$-value of each oscillator.
We have also performed simulations with different initial conditions, such as $(x_j,y_j)=(0.2,-1)$ or $(x_j,y_j)=(-\sqrt{0.5},+\sqrt{0.5})$.
We have not found any noticeable difference in the asymptotic behavior for those initial conditions.
The simulations are performed for cluster sizes $n=50$, $n=25$, $n=10$, $n=5$, $n=2$, and $n=1$.

\begin{figure}[tbp]
$\mathsf{(a)}$\qquad \qquad \qquad \qquad \qquad\qquad\qquad \qquad \qquad\qquad \quad\qquad\qquad\quad\quad\\
\includegraphics[width=\linewidth]{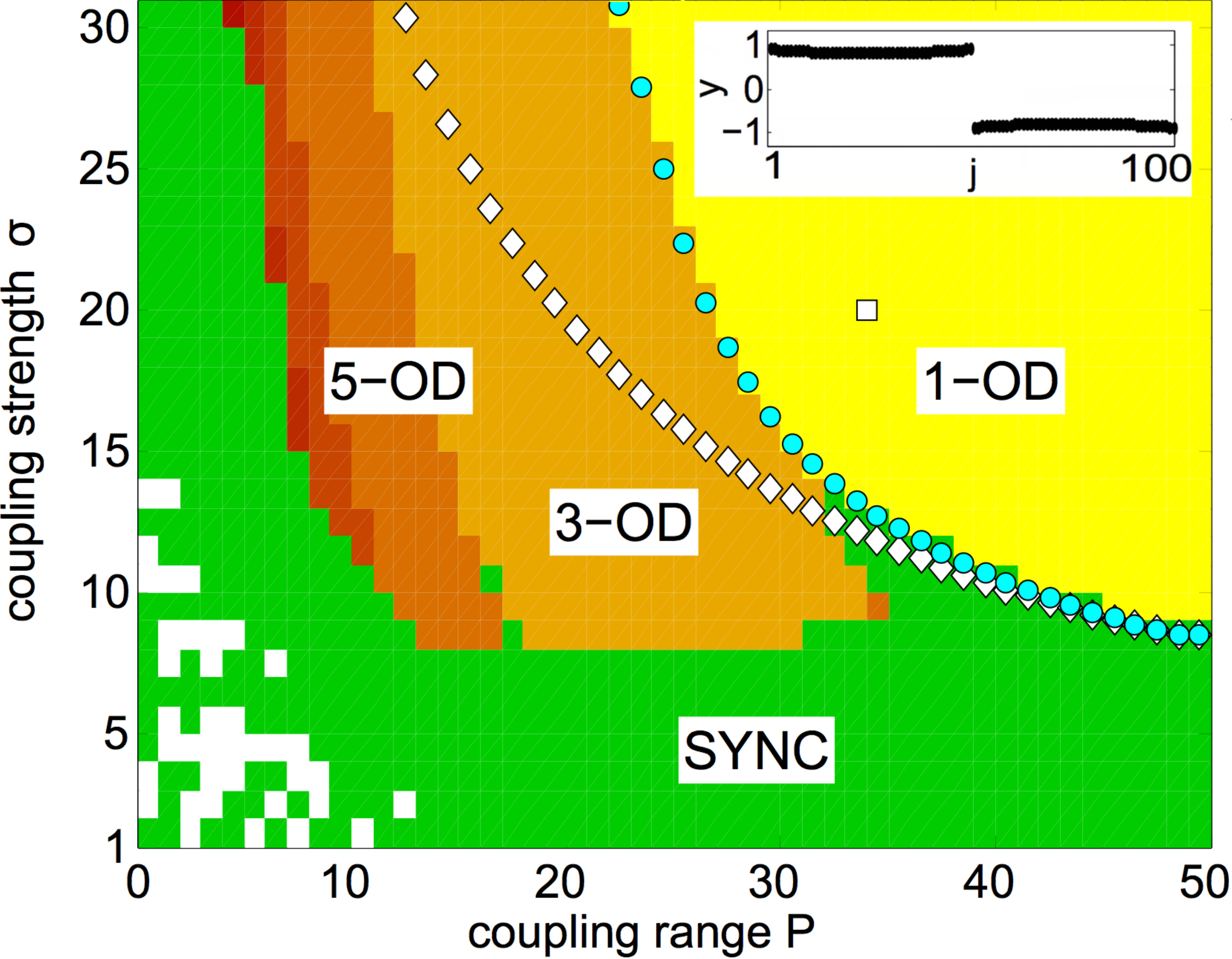}
$\mathsf{(b)}$\qquad \qquad \qquad \qquad \qquad\qquad$\mathsf{(c)}$\qquad \qquad \qquad\qquad \qquad\qquad\qquad\qquad\\
\includegraphics[width=0.49\linewidth]{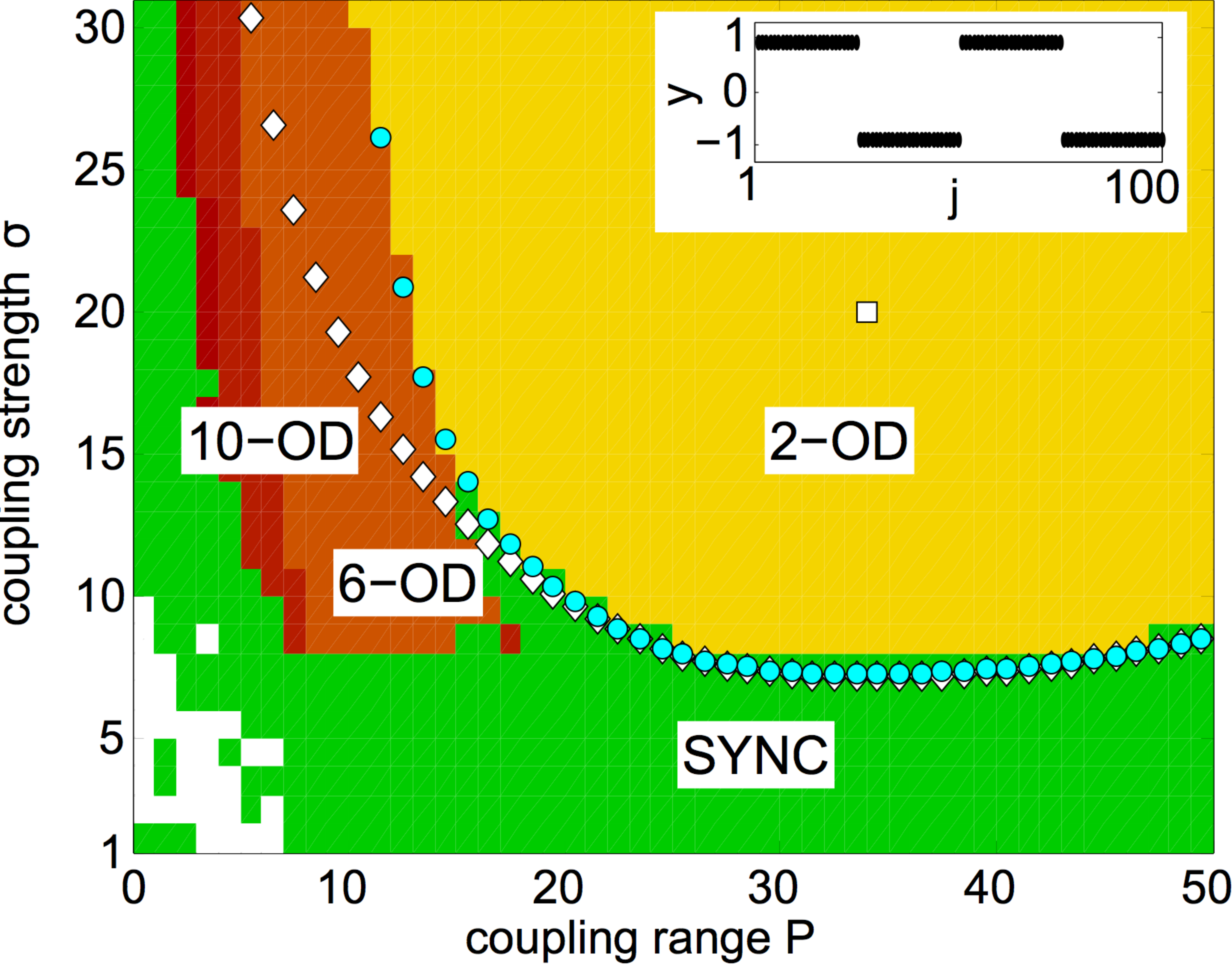}
\includegraphics[width=0.49\linewidth]{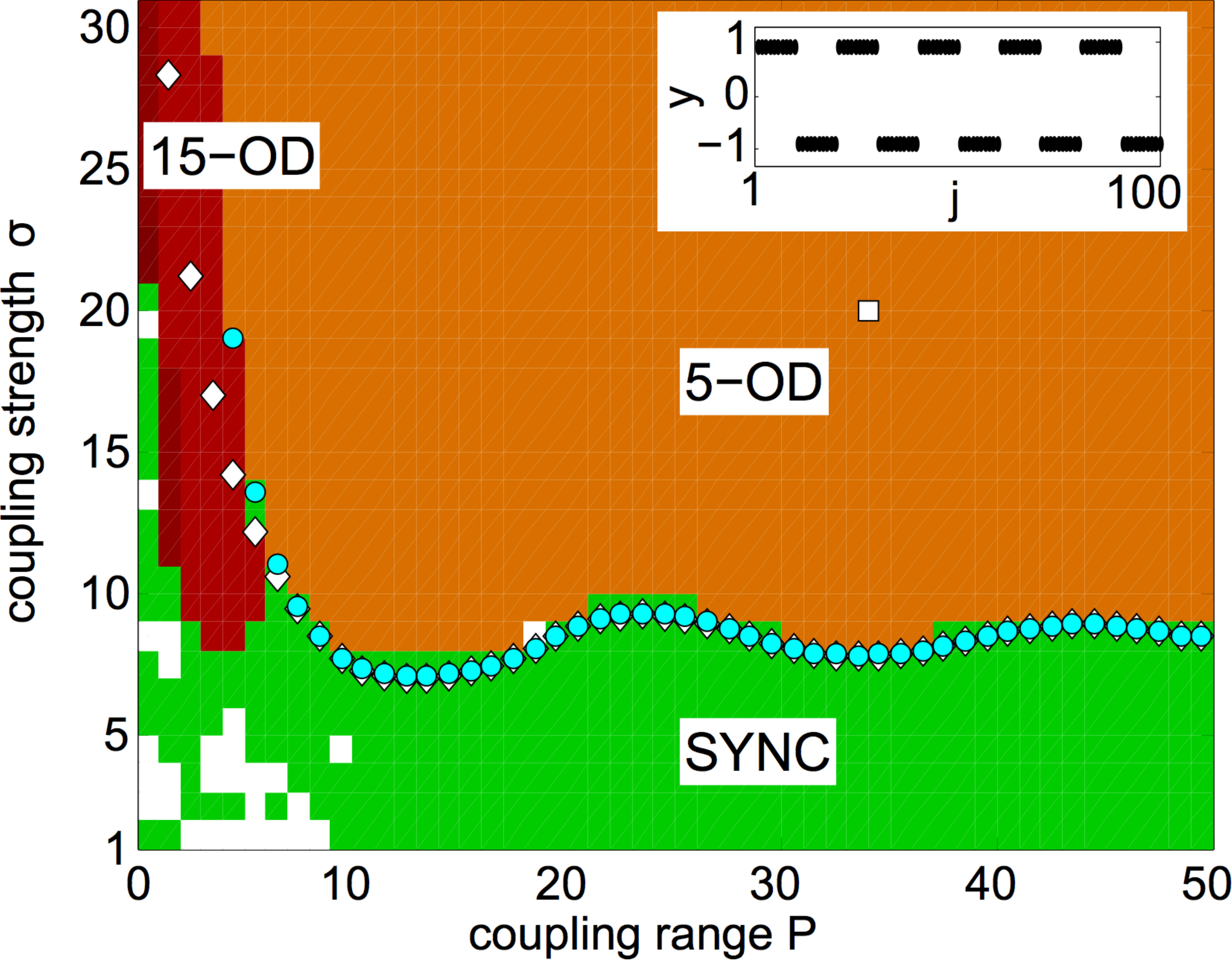}
$\mathsf{(d)}$\qquad \qquad \qquad \qquad \qquad\qquad$\mathsf{(e)}$\qquad \qquad \qquad\qquad \qquad\qquad\qquad\qquad\\
\includegraphics[width=0.49\linewidth]{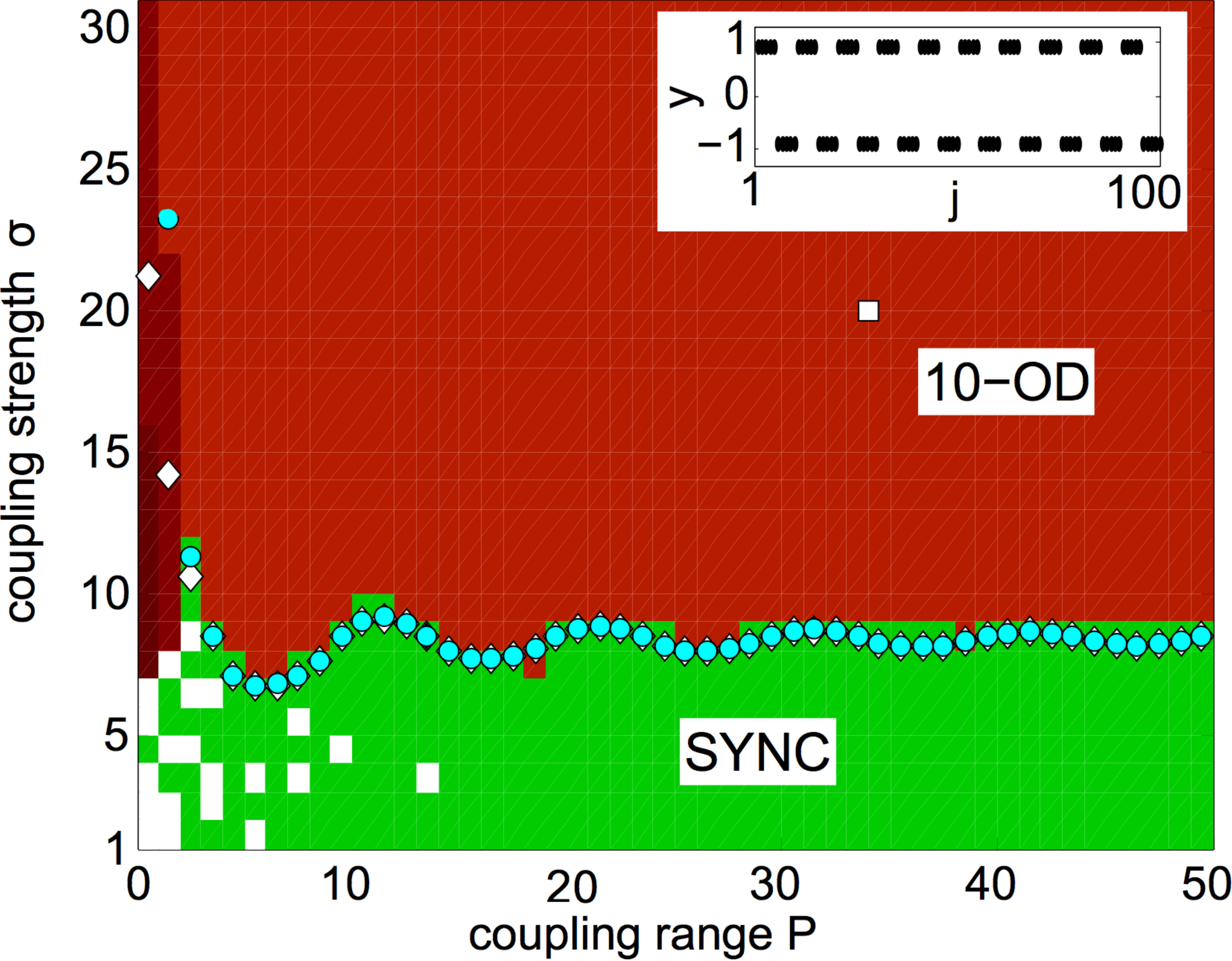}
\includegraphics[width=0.49\linewidth]{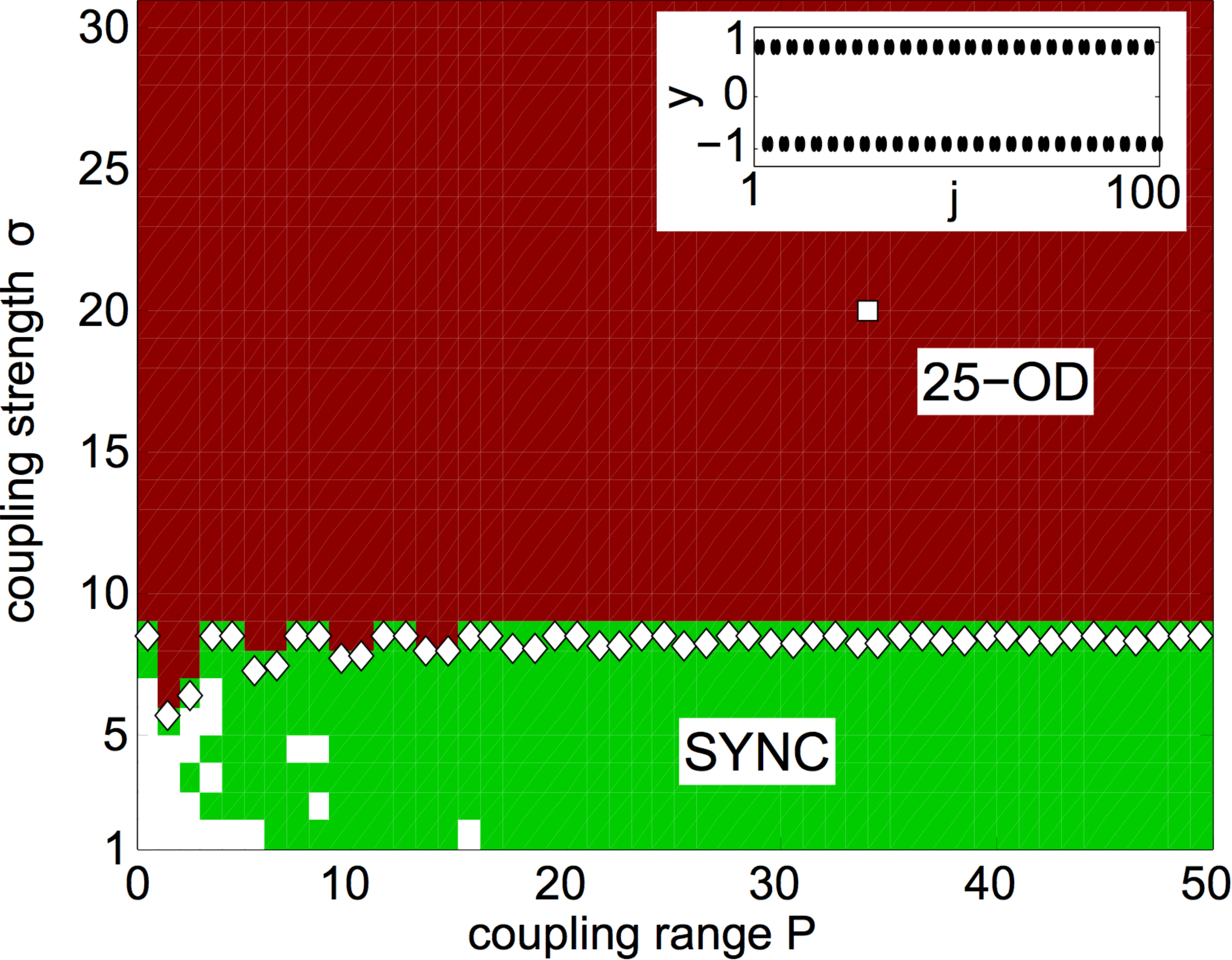}\caption{Numerical simulation of the asymptotic behavior of the system \eqref{eq:StuLaRing}, in dependence on the coupling range $P$ (horizontal axis) and the coupling strength $\sigma$ (vertical axis).
The green (gray) region marks synchronized oscillations, the colors yellow (light) to red (dark) mark the various $m$-cluster oscillation death ($m$-OD) states.  White patches correspond to more complex patterns. The insets show the asymptotic OD state corresponding to $P=35$ and $\sigma=20$ (marked by white squares). Initial conditions:
(a) $n=50$, (b) $n=25$, (c) $n=10$, (d) $n=5$, (e) $n=2$ (cluster size $n=N/(2m)$).
Analytical approximations: White diamonds: mean-field approximation $\sigma_{P,n}$. Light blue circles: beyond mean-field approximation $\sigma^*_{P,n}$. Parameters: $N=100$, $\lambda=1$, $\omega=2$.
}
\label{fig:num2x50}
\end{figure} 

\begin{figure}[tbp]
\begin{center}
\includegraphics[width=\linewidth]{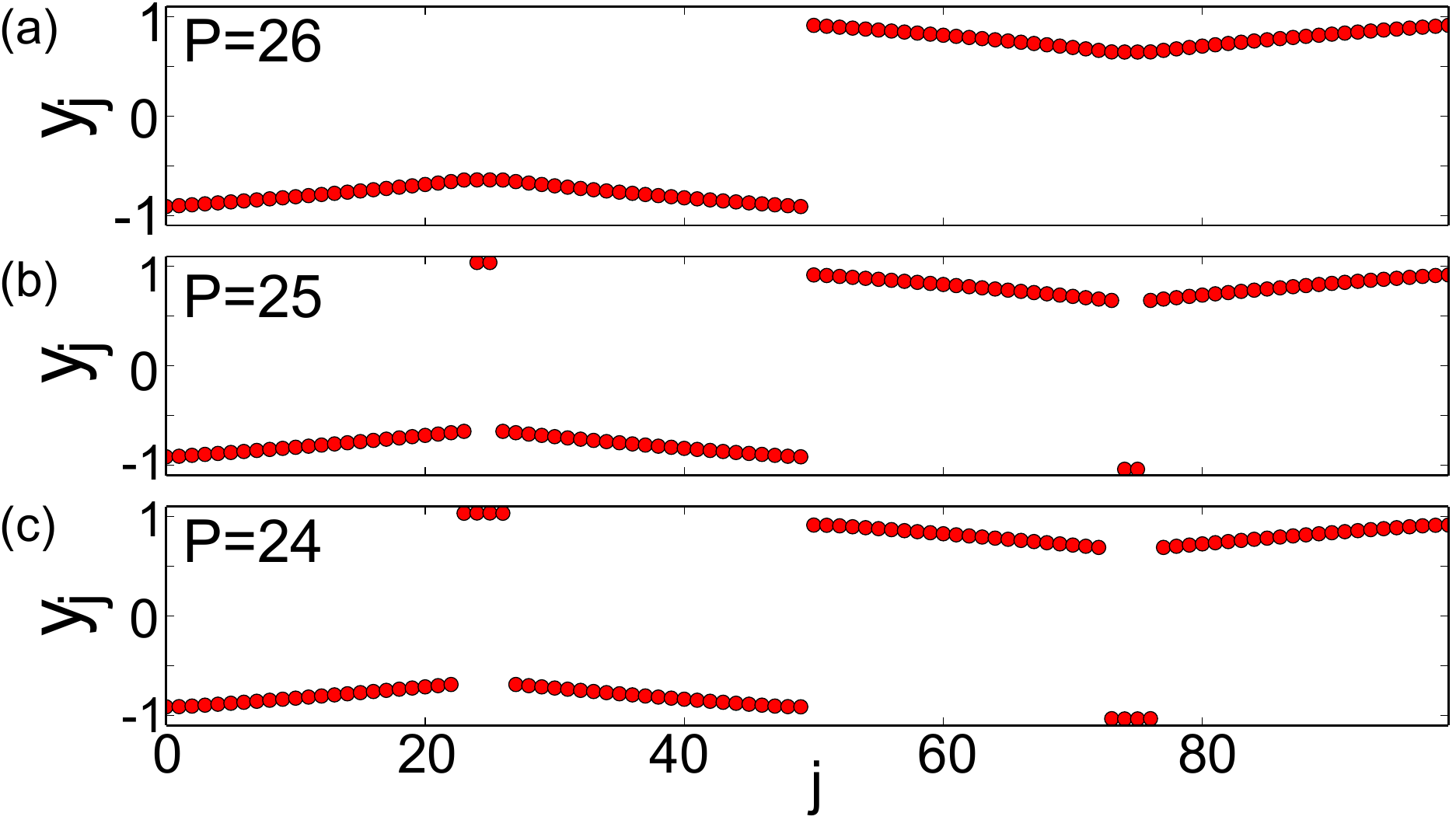}
\caption{Transition from 1-OD to 3-OD. (a) $P=26$. (b) $P=25$. (c) $P=24$. Parameters:  $N=100$, $\lambda=1,\, \omega=2,\, \sigma=25, \, n=50$. The simulations were run up to time $t=500$.}
\label{fig:branches_1OD_3OD}
\end{center}
\end{figure}

The asymptotic results are depicted in Fig.~\ref{fig:num2x50} in the ($P$,$\sigma$)-plane for different initial conditions (panels a - e),
exhibiting high multistability. As discussed above, the initial conditions are regular anti-phase clusters of different size with superimposed random additions. For those initial conditions amplitude chimera states cannot be observed, in contrast to 
\cite{ZAK14,ZAK15b} where other initial conditions were used and also transients were recorded.
We observe a large region of synchronous oscillations for each coupling range $P$. 
We also find regions of stable oscillation death, color-coded by yellow (light), orange (medium), and red (dark), depending on the cluster size. By $m$-cluster oscillation death ($m$-OD), we denote a steady state such that there exist $m$ clusters on the upper branch as well as $m$ clusters on the lower branch.
These clusters do not necessarily have the same size. 
We observe a large region of oscillation death where the clustering takes the same form as given by the initial conditions.
This region can in general be found for large coupling strength $\sigma$ and large coupling range $P/N$.
It is the boundary of this specific region that we investigate analytically below.
Furthermore, we also find regions where there are asymptotically three times as many clusters as in the initial conditions (e.g., 3-OD in panel (a), 6-OD in panel (b), 15-OD in panel (c) of Fig.~\ref{fig:num2x50}), and regions where we find five times as many (e.g., 5-OD in panel (a), 10-OD in panel (b)). 
These regions are marked by darker color. 
A scenario leading from 1-OD to 3-OD is shown in Fig.~\ref{fig:branches_1OD_3OD}. In panel (b) the 1-OD state is unstable, and the two oscillators in the middle of the cluster switch to the opposite branch. 
In panel (c) two more oscillators switch to the opposite branch. 
 
\begin{figure}[tbp]
\includegraphics[width=\linewidth]{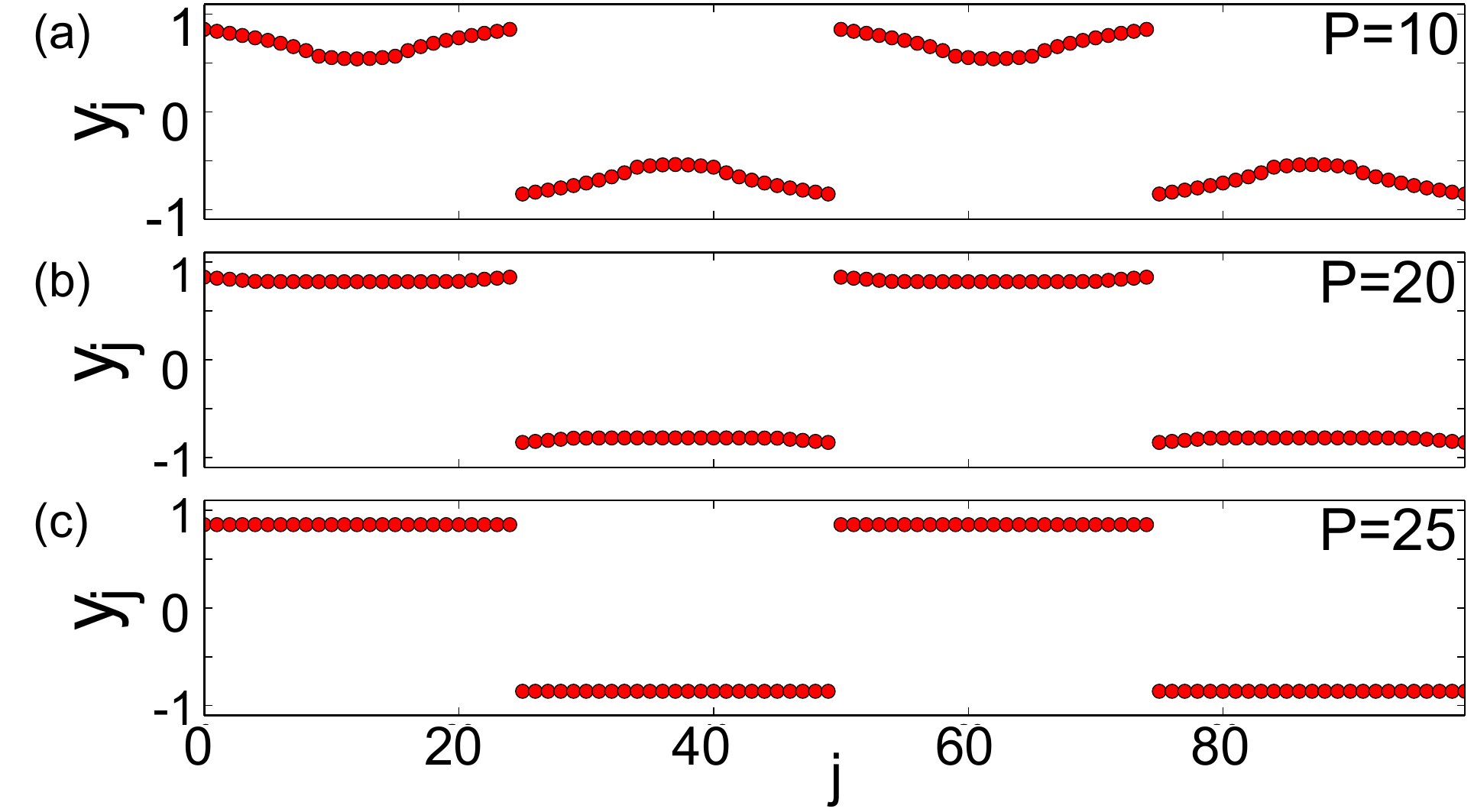}
\caption{Cluster deformation effects of oscillation death for a coupling strength $\sigma = 15$ and different values of the coupling range $P/N$. Snapshots for the variables $y_j$. (a) $P = 10$, (b) $P = 20$, (c) $P  = 25$. Other parameters: $N =100$, $\lambda = 1$, $\omega = 2$, $n = 25$, $m = 2$.}
\label{fig:Branches}
\end{figure}

We observe that, within one cluster, not all oscillators assume the same values for radius and phase: 
We call this effect \emph{cluster deformation} (Fig.~\ref{fig:Branches}).
The deformation is particularly pronounced for small coupling range (Fig.~\ref{fig:Branches}a), and is approximately linear, either decreasing or increasing, seen from the corner oscillators of the cluster.
In the middle of the cluster, we observe a plateau, where the oscillators have the same radius and phase.
We will describe this phenomenon in more detail in section \ref{v}, where it will be used to find a better approximation of the behavior of system \eqref{eq:StuLaRing} beyond mean-field theory.

\begin{figure}[tbp]
\begin{center}
\includegraphics[width=\linewidth]{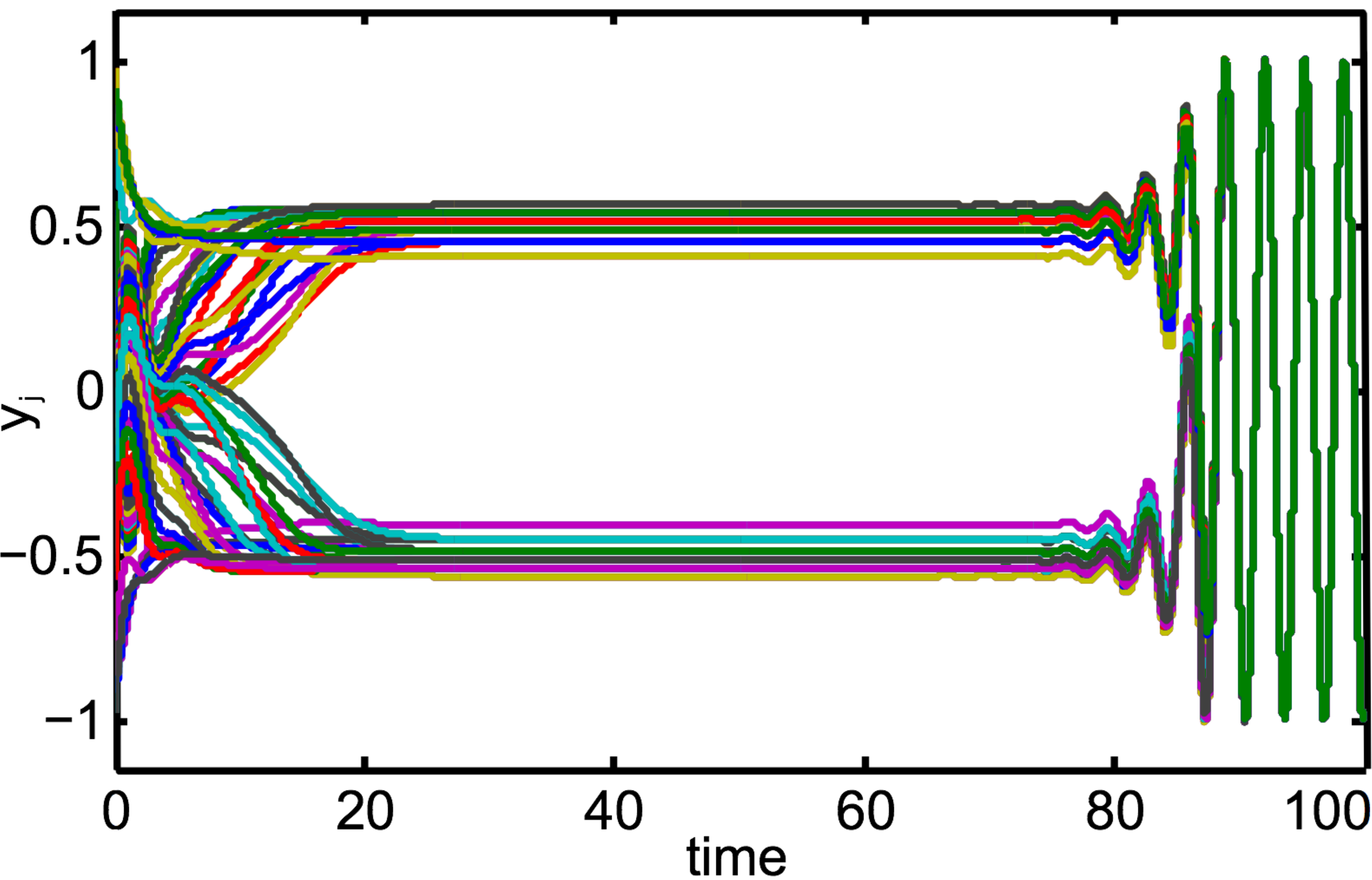}
\caption{Time series $y_j$ of transient oscillation death: the first transition is from the initial condition to an unstable inhomogeneous stationary state that shows a second transition to synchronous oscillations, i.e., \emph{transient oscillation death}. Parameters: $N=100$, $\lambda=1$, $\omega=2$, $\sigma=6$, $P=30$.}
\label{transient}
\end{center}
\end{figure}

\begin{figure}[tbp]
\includegraphics[width=\linewidth]{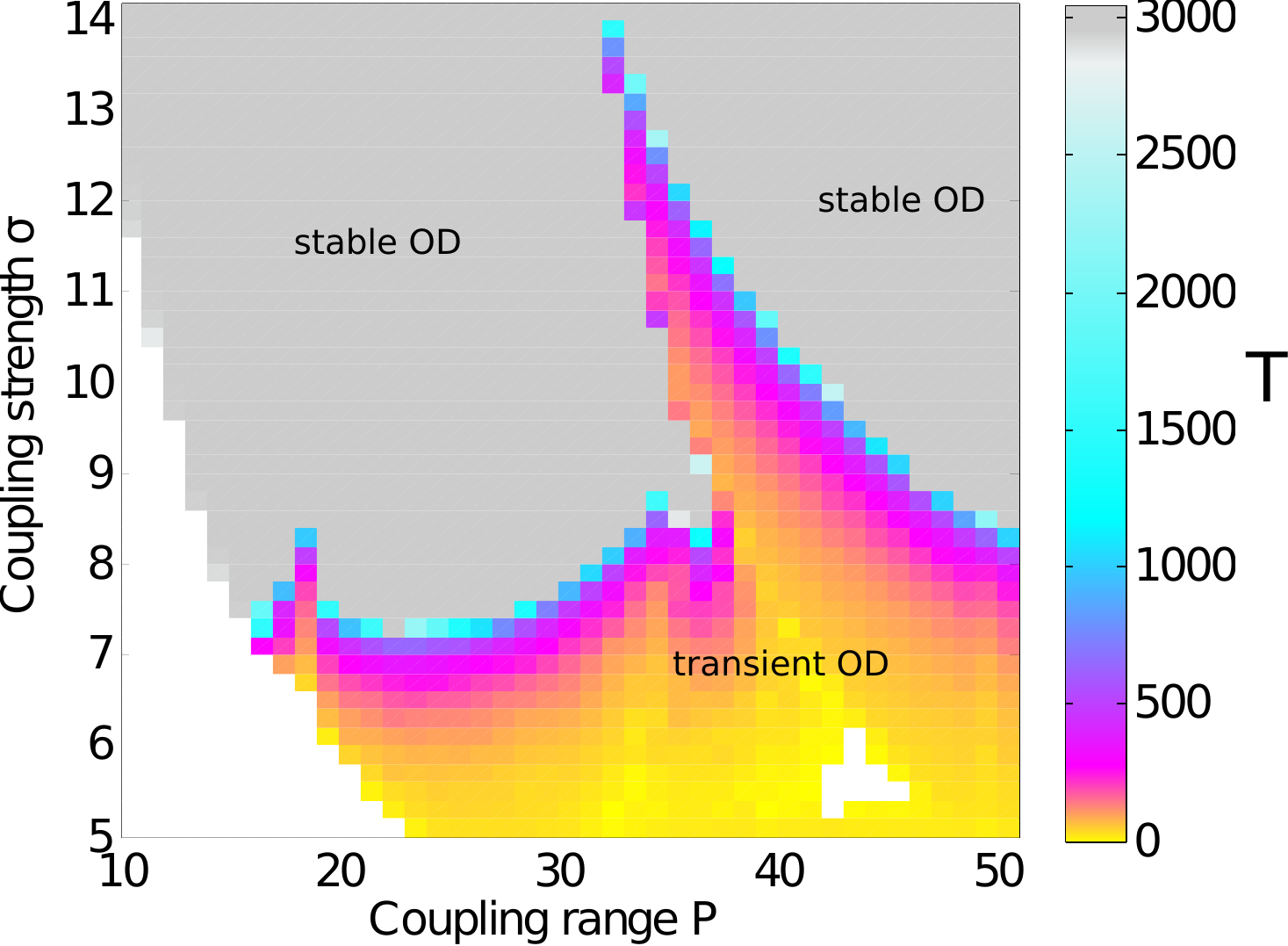}
\caption{Transient time $T$ in the plane of the coupling range $P/N$ and the coupling strength $\sigma$ with 1-OD ($n=50$) initial distribution. The colorcode (grayscale) shows the transient time $T$; the light gray region denotes stable oscillation death. Parameters: $N=100$, $\lambda=1, \, \omega=2$. As initial conditions we used the explicit values given by Eqs.~\eqref{x} and \eqref{y}, but substituted the coupling strength $\sigma$ instead of $\varepsilon$. Here, no additional random numbers were added to the initial conditions.}
\label{fig:transRegi}
\end{figure}

Additionally, in the region where the asymptotic behavior is given by a synchronized oscillation (SYNC), we find transient oscillation death.
Fig.~\ref{transient} depicts the time series of $y_j$, $j=1,\dots,100$, in the transient regime. 
It shows a transition from the initial state to \emph{transient} $3$-OD. 
For a long time, the inhomogeneous steady states seem stable but, eventually, a second transition to synchronous oscillations occurs. 
Following these observations, we have also recorded the transient times $T$, see Fig.~\ref{fig:transRegi}.
We generally observe an increase of $T$ if the coupling strength $\sigma$ is increased.
The transient time will be investigated in more detail in section~\ref{vi}.
 
\section{Mean-field approximation}\label{iv}

It is the purpose of this section to introduce a simplified model, which is able to explain the above numerical observations.
In this simplified model, we assume that we have clusters of size $n$, where the single components of the cluster are supercritical Stuart-Landau oscillators Eq.~\eqref{eq:StuLa}. 
Since we restrict the considerations to cluster distributions with uniform cluster size $n$, we obtain the condition $N=0\,\operatorname{mod}\, 2n$. 
For $N=100$, e.g., this is the case for  $n=50, 25, 10, 5, 2$ and $1$. 
We assume $P$-nearest neighbor coupling, see Eq.~(2).

As a simplifying mean-field assumption, and deviating from the numerics, we assume in this section, and only in this section, that all oscillators on the upper branch have the same radius and phase.
We also assume that they are anti-symmetric (anti-phase) with respect to the oscillators on the lower branch, implying that those also all have the same radius and phase.
Therefore the coupling term vanishes within the clusters.
It is between the clusters that the coupling term gives non-zero contributions (coupling of the real parts, inducing the inhomogeneous steady states, cf. section~\ref{ii}).

Our goal is to replace the model of $N$ coupled oscillators by a simplified model of \emph{two Stuart-Landau oscillators} which are coupled via their real parts with an \emph{effective coupling strength} $ \kappa_{P,n} \sigma$:
\begin{equation}
\begin{split}
\dot{z}_{1} &= f(z_1)+\kappa_{P,n} \sigma \left(\mbox{Re}\, z_{2}-\mbox{Re}\, z_1\right)\\
\dot{z}_{2} &= f(z_2)+ \kappa_{P,n} \sigma \left(\mbox{Re}\, z_{1}-\mbox{Re}\, z_2\right)
\end{split}
\end{equation}
Since the dynamics of this system can be described analytically, cf. section~\ref{ii}, we can draw conclusions about the onset of oscillation death, provided that we know the effective coupling strength $\kappa_{P,n} \sigma$.
Our task hence is to find the scaling factor $\kappa_{P,n}$, depending on both the coupling range $P$ and the cluster size $n$, as indicated by the numerical simulations.

As a first reduction step, note that it is sufficient to consider only the links emanating from one cluster instead of those of the complete network.
Indeed the network structure implies that the coupling structure is the same for every cluster. 
This reduces the number of oscillators for which we have to calculate the effective coupling from $N$ to $n$.

\begin{figure}[tbp]
\begin{center}
\includegraphics[width=\linewidth]{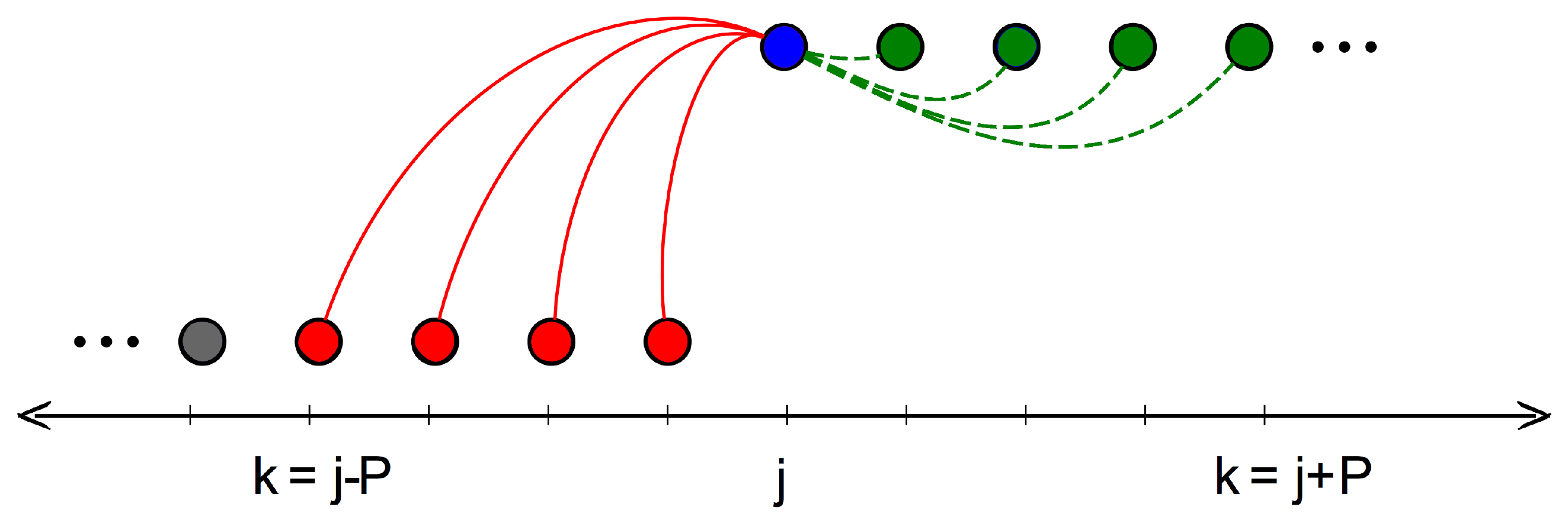}
\caption{Scheme of nonlocal coupling. For the oscillator $j$ and $P=4$ we have 4 relevant links to the left (solid red) as well as 4 irrelevant links to the right (dashed green). Irrelevant links do not contribute to the coupling term.
}
\label{fig:coupling_scheme}
\end{center}
\end{figure}

In the following, we number the oscillators within one cluster with indices $j=0, \dots, n-1$, where $n$ is the size of the cluster. 
The oscillator with index $j$ is coupled to $2P$ neighbors: $P$ to the left, and $P$ to the right. We count \emph{`relevant'} links $(+1)$, which are those between oscillators of different clusters. 
In contrast, \emph{`irrelevant'} links, which are the links between oscillators of the same cluster, are not counted $(0)$; see Fig.~\ref{fig:coupling_scheme}. 

As a first step, consider a coupling range $P<2n$. For the oscillator $j$ we now count $P$ links to the left. Of these, a maximum of $P-j$ is relevant, since the first $j$ links to the left are between oscillators of the same cluster. If  $P-j$ is negative, then all left links are irrelevant.
The number of relevant left links is therefore bounded from below by zero.
The upper bound of the  number of relevant links to the left is $n$, corresponding to the full cluster being relevant.
In conclusion, the number of relevant links to the left of oscillator number $j$ is given by 
\begin{equation}
\max\big\{\min\left\{n,P-j\right\},0\big\}.
\end{equation}

Now also consider coupling ranges $P>2n$, more precisely, consider $P \in \{2nk + 1, \dots , 2n(k+1)\}$, $k \in \mathbb{N}_0$.
Here $k=0$ corresponds to $P \le 2n$. 
Note that in this case the respective lower bound of relevant links to the left is given by $kn$ ( i.e.~$k$ full clusters), while the upper bound is given by $(k+1)\,n$ (i.e.~$(k+1)$ full clusters).
Consequently, for the oscillator $j$, we find that the number of relevant links to the left is given by 
\begin{equation}
\max\big\{\min\left\{(k+1)\,n,P-j-kn\right\},kn\big\}.
\end{equation}
By reflection, the number of relevant links to the left of oscillator $j$ is the same as the number of relevant links to the right of oscillator $n-j-1$.

In the next step, in order to reduce from $n$ oscillators to two oscillators, we calculate the \emph{mean relevant coupling} of the whole cluster, and therefore consider each cluster as lumped into only one effective oscillator.
We sum up the respective value of each oscillator within the cluster,  and divide by the number of neighbors $P$ as well as by the number of oscillators $n$ to obtain a first approximation for the average coupling $\kappa_{P,n}$:
\begin{equation}
\kappa_{P,n} = \frac{1}{Pn}\sum\limits_{j=0}^{n-1}\max\big\{\min\left\{(k+1)\,n,P-j-kn\right\},kn\big\}
\end{equation}
This can be inserted into the threshold condition for stable oscillation death obtained in section~\ref{ii}, $\varepsilon_{H\!B}=\frac{1}{4}(\lambda+4\omega^2/\lambda)$.
In total, for $P \in \left\{2kn+1,\dots,2\left(k+1\right)n\right\}$, $k \in \mathbb{N}_0$, we find the threshold
\begin{equation}
\sigma_{P,n}=\frac{\tfrac{1}{4}\left(\lambda+4 \omega^2 /\lambda \right) P n}{\sum_{j=0}^{n-1} \max\big\{\min\left\{(k+1)n,P-j-kn\right\},kn\big\}}
\end{equation}
for the onset of oscillation death.

Note that the size of the system, i.e., the total number of oscillators $N$, does not appear in the threshold.

We can compare this analytic mean-field result with the numerical simulations:
For a coupling range $P$ which is of the same order of magnitude as the cluster size $n$, or larger, the analytic and the numeric thresholds agree very well.
For a coupling range $P$ which is small compared to the cluster size, we note that the analytic results give smaller thresholds than the simulations, see white diamonds in Fig.~\ref{fig:num2x50}.
This suggests that the simplified mean-field model must be improved for this parameter regime.

\section{Beyond mean-field theory} \label{v}

As mentioned above, the assumption that all oscillators within a cluster have the same radius and phase is not true in general and deviates from the numerical results. 
Therefore, we aim for a correction term to the mean-field theory used above, depending on the coupling range $P$ and the cluster size $n$.

In our reduced model, we replace the mean-field coupling $\kappa_{P,n} \sigma$ by the corrected mean coupling $\left(\kappa_{P,n}-\kappa^*_{P,n}\right) \sigma$, and it is the aim of this section to estimate the correction term $\kappa^*_{P,n}$.

In Fig.~\ref{fig:Branches} we observed a deformation of the cluster shape (branch splitting) for small $P$.
The deformation is approximately linear, where the $x$- and the $y$-variable of the two corner oscillators of the cluster stay (approximately) the same. 
The values of the neighboring oscillators, proceeding inwards, either decrease or increase.
Seen from the left, there are 
$\min\{P \,\mbox{mod}\, n, (P+1) \,\mbox{mod}\, n, n-(P+1)\,\mbox{mod}\,n,n-P\,\mbox{mod}\,n \}$ oscillators which show a linear decrease/increase with respect to their left neighbor.
To first order, it follows that there exists a plateau, where the oscillators have the same radius and phase.
On the right we encounter an analogous linear increase or decrease, respectively.
The cluster is reflection symmetric in the sense that the phenomenon is the same if we approach from the left or the right.

This behavior is due to the fact that not all oscillators ``feel" the same coupling:
The corner oscillators feel more or less oscillators from the other clusters than those in the middle, depending on the precise values of $P$ and $n$.

An exception are the values $P\in \{ n k, \, nk-1: k \in \mathbb{N}\}$  where the mean coupling coincides exactly with the coupling that every single oscillator feels; thus no branch splitting effects occur (Fig.~\ref{fig:Branches}c).
In particular, we can conclude that, for those $P$, no correction term is necessary.

Following the description above, the effect of the cluster deformation on the mean coupling is proportional to $\min\{P \,\mbox{mod}\, n, (P+1) \,\mbox{mod}\, n, n-(P+1)\,\mbox{mod}\,n,n-P\,\mbox{mod}\,n \}$ as well as proportional to the cluster size $n$.
Overall, the significance of the correction term decreases as $P$ increases, since the more oscillators are coupled, the closer we are to the mean-field approximation. A simple intuitive ansatz is to assume a global $1/P$ dependence of the correction term. Thus, to compensate the cluster deformation term, we need a prefactor $P^{-2}$ to achieve the desired global $1/P$ dependence.

We suggest the following term:
\begin{align}
\kappa^*_{P,n}=c  nP^{-2} \min\{ & P\, \mbox{mod}\, n, (P+1)\, \mbox{mod}\, n, \nonumber \\
     & n-(P+1)\,\mbox{mod}\, n,n-P\,\mbox{mod}\, n \}.
\end{align}
Here, we denote the proportionality constant by $c$.
We have found it to be the same for all our examples, $c\approx0.047$.
We therefore assume that this constant is independent of $n$ and $P$.
Furthermore, we also claim that it is independent of $N$, since the single oscillators are not influenced by the circumference of the ring. 
However, there might be an additional dependence on $\lambda$ and $\omega$, which is not the subject of our present
investigation.

As a consequence, we also obtain a corrected formula for the oscillation death threshold, which we denote by $\sigma^*_{P,n}$,
\begin{equation}\label{OD-thresh}
 \sigma^*_{P,n}=\frac{\lambda+4 \omega^2/\lambda}{4 \big(\kappa_{P,n}-\kappa^*_{P,n}\big)}.
\end{equation}
Note that Eq.~(\ref{OD-thresh}) gives either a larger threshold or the same threshold as before.
The difference of the simplified and the improved formula is most noticeable if $P$ is small compared to the cluster size $n$, compare Fig.~\ref{fig:num2x50}.
The improved formula gives excellent agreement with the numerical threshold for stable oscillation death. 

\section{Transient behavior}\label{vi}

The aim of this section is to describe the transient times analytically.
Transient oscillation death, see e.g.~the time series depicted in Fig.~\ref{transient}, occurs for coupling strengths $\sigma$ between the emergence of the unstable inhomogeneous steady state branches (via a pitchfork bifurcation, cf. section~\ref{ii} and \cite{ZAK13a}) and the stabilizing Hopf bifurcation. 
The manifestation of the long transient times is due to different orders of magnitude of the real parts of the eigenvalues 
\begin{align}
\mbox{Re}\,\mu_{1,2} &= -\lambda + 2\bar{\kappa}_{P,n}\sigma-2\sqrt{\bar{\kappa}_{P,n}^2\sigma^2-\omega^2}  \\
\mbox{Re}\,\mu_3 &= \mu_3=-2\sqrt{\bar{\kappa}_{P,n}^{2}\sigma^2-\omega^2}.
\end{align}
Here we have used the abbreviation $\bar{\kappa}_{P,n}=\kappa_{P,n}-\kappa^*_{P,n}$.
The absolute value of $\mu_3$ increases with increasing coupling, see also Fig.~\ref{triplett} for the system of two coupled oscillators.
Note that the real part of the complex conjugate eigenvalues $\mu_{1,2}$ is positive and smaller than $1$ and in fact slowly goes to $0$ as $\sigma$ approaches the stabilizing Hopf bifurcation, see Fig.~\ref{triplett}, where $\mbox{Re}\,\mu_1$ is depicted in green (dashed) and crosses the real axis at the Hopf bifurcation point.

\begin{figure}[htbp]
\includegraphics[width=\linewidth]{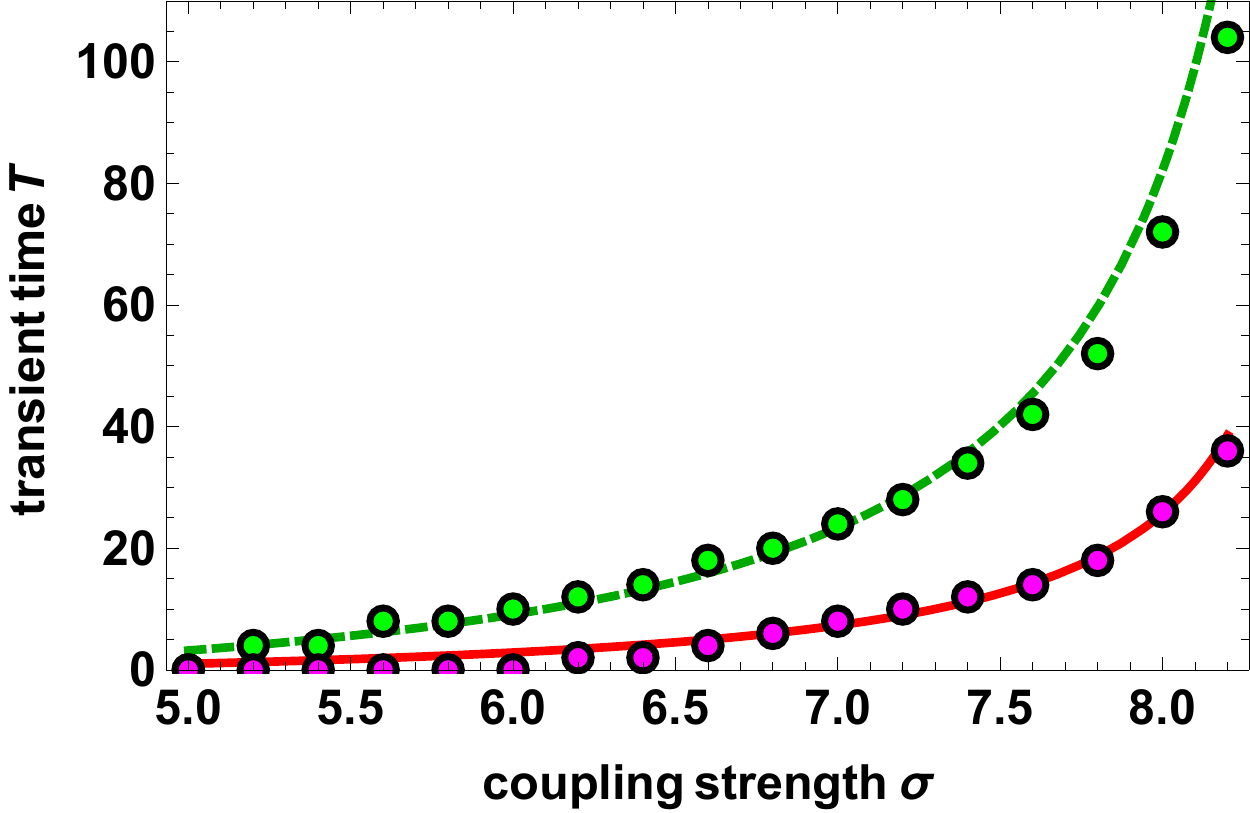}
\caption{Transient time $T$ vs. coupling strength $\sigma$ for different variances of the initial distribution for 1-OD. Initial conditions $(x_j,y_j)=(-0.2,+0.7)$ on the upper branch, and $(x_j,y_j)=(+0.2,-0.7)$ on the lower branch, with random fluctuations of different variance added. Green (lighter) dots: Numerical values using a variance of $10^{-8}$. The dashed green line gives the analytical values calculated in Eq.~\eqref{trans} with fit parameter $\beta=1.22$. Magenta (darker) dots: Numerical values using a variance of $10^{-4}$. The solid red line gives the analytical values calculated in Eq.~\eqref{trans} with  $\beta=0.382$.
Parameters: $N=100$, $\lambda=1$, $\omega=2$, $\bar{\kappa}_{P,n}=0.48898$,  $P=48$, $n=50$, $m=1$. }
\label{comparison}
 \end{figure}

The (slow) unstable eigenvalue $\mu_1$  is compensated by the (fast) stable eigenvalue $\mu_3$.
The transient time $T \sim 1/\mbox{Re}\,\mu_1$ is shorter if the real part of $\mu_1$ increases.
If $\mbox{Re}\,\mu_1$ goes to zero (which it does at the Hopf bifurcation point), the transient time goes to infinity.
In contrast, the transient time increases with the absolute value of the stable eigenvalue.

We considered an additive Gaussian randomization of initial data, with fixed variance. 
Numerical experiments then indicated that the observed mean transient time is proportional to the ratio of the real parts of the eigenvalues, more precisely, the larger the ratio, the longer the transient time. 
The stochastic dependence of $T$ on the initial conditions, i.e., the position and variance of the distribution of the oscillators will not be pursued in this publication.
We include this by a factor $\beta>0$, which so far we have determined numerically, by fitting one or more values of $\sigma$.
Here $\beta$ depends only on the initial conditions and is independent of $P, \, n, \, N$ and $\sigma$. 
 
We conclude that the dependence on $P$ and $\sigma$ of the expection value $\langle T\rangle$ of the transient time $T$ can be described by
\begin{align}
\langle T\rangle&=\beta\,|\mbox{Re} \, \mu_3|/\mbox{Re}\,\mu_1\\
&=\beta \frac{ 2\sqrt{\bar{\kappa}_{P,n}^{2}\sigma^2-\omega^2}}{ -\lambda+2 \bar{\kappa}_{P,n}\sigma-2\sqrt{\bar{\kappa}_{P,n}^{2}\sigma^2-\omega^2}}. \label{trans}
\end{align}
Note that, due to the sensitive dependance on the initial conditions, we only give an estimate for the expectation value of the transient time $T$.
Fig.~\ref{comparison} shows the numerically calculated transient times vs $\sigma$ for different variances of the initial distribution.
The solid curves represent Eq.~\eqref{trans} with fit parameters $\beta$.

\section{Conclusion} \label{vii}

We have investigated a network of nonlocally coupled Stuart-Landau oscillators under a coupling which breaks $S^1$-symmetry.
Following numerical simulations with clustered initial conditions, we have found a family of oscillation death states (inhomogeneous steady states) which can be distinguished by their cluster size. Clustering always emerges in pairs of in-phase and anti-phase clusters.
Two different cases occur:
Either the pattern of the initial cluster configuration coincides with the asymptotic cluster distributions, or additional clusters appear in the middle of the original clusters.
The shape of the clusters depends on the coupling range $P$, i.e., the number of nearest neighbors which are coupled.
If $P$ is either an integer multiple of the cluster size, or an integer multiple minus one, then all oscillators within the cluster exhibit the same radius and phase.
In all other cases a linear cluster deformation (branch splitting) occurs.
We have developed an approximate analytical description by a reduced model of two mean-field coupled Stuart-Landau oscillators.
In particular, we have analytically calculated the onset of stable oscillation death with prescribed clustering.
Our theory goes beyond standard mean-field theory, which only gives a rough estimate.
Specifically, we have extended the mean-field coupling by an approximation for the spatial deviation from the mean.
As a consequence, we are able to predict the boundaries of the different stability regimes of $m$-cluster oscillation death analytically with high precision.

In addition to the asymptotically stable oscillation death states, we also find a region where oscillation death is transient: 
It persists for a long time but then disappears in favor of synchronized oscillations.
The transient behavior occurs due to the interaction of a slow unstable eigenvalue and a fast stable eigenvalue of the inhomogeneous steady state.
We have analytically calculated a scenario leading from the transient behavior to the asymptotically stable oscillation death with the coupling strength serving as a bifurcation parameter.
Near the stabilizing Hopf bifurcation, the calculated transient time goes to infinity which is in accordance with the numerical simulations.

Note that in this paper we have confined attention to spatially coherent clusters of inhomogeneous steady states. As shown in \cite{ZAK14,ZAK15b}, there also exist hybrid states consisting of coexisting domains of spatially coherent and incoherent oscillation death, called {\em chimera death}, for certain initial conditions.
This is an indication of the high multistability of the system. 
The regimes of existence of 1-cluster, 3-cluster, 5-cluster etc. chimera death in the ($\sigma$, $P$) parameter plane is similar to Fig.~\ref{fig:num2x50}. 
In addition, for small $P$ there exist transient amplitude chimeras of coexisting domains of spatially coherent (synchronized) and incoherent oscillations.

Stuart-Landau oscillators as normal form of systems near Hopf bifurcation can serve as a model system for many applications.
The set-up for both the mean-field theory and its correction does not depend on the specific form of Stuart-Landau oscillators.
Similar symmetric cluster states, maybe even including more branches, can be expected in any network with a clearly defined 
symmetry, in which case the methods presented in this paper can be employed.
\section{Acknowledgments}
This work was supported by SFB 910 ``Control of Self-Organizing Nonlinear
Systems: Theoretical Methods and Concepts of Application" of the Deutsche Forschungsgemeinschaft.

The authors would like to thank Matthias Bosewitz  for helpful comments and discussions.


\begin{thebibliography}{10}

\bibitem{PIK01}
A. Pikovsky, M.~G. Rosenblum, and J. Kurths, {\em Synchronization, A Universal
  Concept in Nonlinear Sciences} (Cambridge University Press, Cambridge, 2001).

\bibitem{BOC02}
S. Boccaletti, J. Kurths, G. Osipov, D.~L. Valladares, and C.~S. Zhou, Phys.
  Rep. {\bf 366},  1  (2002).

\bibitem{DAH12}
T. Dahms, J. Lehnert, and E. Sch{\"o}ll, Phys. Rev.~E {\bf 86},  016202
  (2012).

\bibitem{DO12}
A.~L. Do, J.~M. H\"ofener, and T. Gross, New J. of Phys. {\bf 14},  115022
  (2012).

\bibitem{PEC14}
L.~M. Pecora, F. Sorrentino, A.~M. Hagerstrom, T.~E. Murphy, and R. Roy, Nat.
  Commun. {\bf 5},  4079  (2014).

\bibitem{POE15}
W. Poel, A. Zakharova, and E. Sch{\"o}ll, Phys. Rev.~E {\bf 91},  022915
  (2015).

\bibitem{FIE09}
B. Fiedler, V. Flunkert, P. H{\"o}vel, and E. Sch{\"o}ll, Phil. Trans.~R.
  Soc.~A {\bf 368},  319  (2010).

\bibitem{SCH13b}
I. Schneider, Phil. Trans.~R. Soc.~A {\bf 371},  20120472  (2013).

\bibitem{SCH14d}
I. Schneider and M. Bosewitz, Discret. Contin. Dyn.~S.~A {\bf 36},  451  (2016).

\bibitem{KUR02a}
Y. Kuramoto and D. Battogtokh, Nonlin. Phen. in Complex Sys. {\bf 5},  380
  (2002).

\bibitem{ABR04}
D.~M. Abrams and S.~H. Strogatz, Phys.~Rev.~Lett. {\bf 93},  174102  (2004).

\bibitem{PAN15}
M.~J. Panaggio and D.~M. Abrams, Nonlinearity {\bf 28},  R67  (2015).

\bibitem{ZOU09a}
W. Zou, X.~G. Wang, Q. Zhao, and M. Zhan, Front. Phys. China {\bf 4},  97
  (2009).

\bibitem{PRA10}
A. Prasad, M. Dhamala, B.~M. Adhikari, and R. Ramaswamy, Phys. Rev. E {\bf 81},
   027201  (2010).

\bibitem{KOS10}
A. Koseska and J. Kurths, Chaos {\bf 20},  045111  (2010).

\bibitem{KOS13}
A. Koseska, E. Volkov, and J. Kurths, Phys. Rep. {\bf 531},  173  (2013).

\bibitem{ZAK14}
A. Zakharova, M. Kapeller, and E. Sch{\"o}ll, Phys.~Rev.~Lett. {\bf 112},
  154101  (2014).

\bibitem{ZAK15b}
A. Zakharova, M. Kapeller, and E. Sch{\"o}ll, J. Phys. Conf. Series, arXiv
  (2015), 1503.03371.

\bibitem{BAN15}
T. {B}anerjee, ArXiv:1409.7895 (2015)

\bibitem{PRE15}
K. Premalatha, V.~K. Chandrasekar, M. Senthilvelan, and M. Lakshmanan,
  ArXiv:1503.07004  (2015).

\bibitem{DOL88}
M. Dolnik and M. Marek, J. Phys. Chem. {\bf 92},  2452  (1988).

\bibitem{CRO89}
M.~F. Crowley and I.~R. Epstein, J. Phys. Chem. {\bf 93},  2496  (1989).

\bibitem{TOI08}
M. Toiya, V.~K. Vanag, and I.~R. Epstein, Angew. Chem. Int. Ed. {\bf 47},  7753
   (2008).

\bibitem{HEI10}
M. Heinrich, T. Dahms, V. Flunkert, S.~W. Teitsworth, and E. Sch{\"o}ll,
  New~J.~Phys. {\bf 12},  113030  (2010).

\bibitem{ZEY01}
K.~P. Zeyer, M. Mangold, and E.~D. Gilles, J. Phys. Chem. A {\bf 105},  7216
  (2001).

\bibitem{CUR10}
R. Curtu, Phys. D {\bf 239},  504  (2010).

\bibitem{KOS10a}
A. Koseska, E. Volkov, and J. Kurths, Chaos {\bf 20},  023132  (2010).

\bibitem{TSA06}
K. Tsaneva-Atanasova, C.~L. Zimliki, R. Bertram, and A. Sherman, Biophys. J.
  {\bf 90},  3434  (2006).

\bibitem{SUZ11}
N. Suzuki, C. Furusawa, and K. Kaneko, PLoS ONE {\bf 6},  e27232  (2011).

\bibitem{ZAK13a}
A. Zakharova, I. Schneider, Y.~N. Kyrychko, K.~B. Blyuss, A. Koseska, B.
  Fiedler, and E. Sch{\"o}ll, Europhys. Lett. {\bf 104},  50004  (2013).

\bibitem{HEN13}
C.~R. Hens, O.~I. Olusola, P. Pal, and S.~K. Dana, Phys. Rev. E {\bf 88},
  034902  (2013).
  
\bibitem{NAN14}   
M. Nandan, C.~R. Hens, P. Pal, and S.~K. Dana, Chaos {\bf 24}, 043103 (2014).

\bibitem{LIU12c}
W. Liu, E. Volkov, J. Xiao, W. Zou, and M. Zhan, Chaos {\bf 22},  033144
  (2012).

\bibitem{ZOU13}
W. Zou, D.~V. Senthilkumar, M. Zhan, and J. Kurths, Phys. Rev. Lett. {\bf 111},
   014101  (2013).

\bibitem{KAR14}
R. Karnatak, R. Ramaswamy, and U. Feudel, Chaos, Solitons and Fractals {\bf 68}, 48 (2014).

\end{thebibliography}


\end{document}